\documentclass[10pt, a4paper, conference]{IEEEtran}
\usepackage{cite}
\usepackage{amsmath,amssymb,amsfonts}
\usepackage{mathrsfs}
\usepackage[ruled,boxed,noend, linesnumbered]{algorithm2e}
\usepackage{graphicx}
\usepackage{textcomp}
\usepackage{bbm}
\usepackage{xcolor}
\usepackage{stfloats}
\usepackage{bm}
\usepackage{amsmath, bm}
\usepackage{subfigure}
\usepackage{epstopdf}
\usepackage{flushend}
\usepackage{array}
\usepackage{booktabs}

\usepackage{multirow}
\usepackage{amsthm}
\usepackage{makecell}
\usepackage{caption}

\graphicspath{{./figs/}}

\allowdisplaybreaks[4]

\def\BibTeX{{\rm B\kern-.05em{\sc i\kern-.025em b}\kern-.08em
  T\kern-.1667em\lower.7ex\hbox{E}\kern-.125emX}}

\DeclareMathOperator*{\argmin}{arg\,min}
\SetKwRepeat{Do}{do}{while}

\SetCommentSty{mycommfont}

\makeatletter

\newcommand{\Rmnum}[1]{\expandafter\@slowromancap\romannumeral #1@}
\makeatother

\SetKwInput{KwInput}{Inputs}
\SetKwInput{KwReturn}{ruturn}
\SetKwInput{KwInit}{Initialization}

\begin{document}

\title{Modular Foundation Model Inference at the Edge:\\ Network-Aware Microservice Optimization}
\author{\IEEEauthorblockN{Juan Zhu, Zixin Wang, Shenghui Song, Jun Zhang, and Khaled B. Letaief}
\IEEEauthorblockA{Dept. ECE., The Hong Kong University of Science and Technology, Hong Kong}
\IEEEauthorblockA{Email: jzhucu@connect.ust.hk, \{eewangzx, eeshsong, eejzhang, eekhaled\}@ust.hk}
}

\maketitle
\begin{abstract}
Foundation models (FMs) unlock unprecedented multimodal and multitask intelligence, yet their cloud-centric deployment precludes real-time responsiveness and compromises user privacy. Meanwhile, monolithic execution at the edge remains infeasible under stringent resource limits and uncertain network dynamics. To bridge this gap, we propose a microservice-based FM inference framework that exploits the intrinsic functional asymmetry between heavyweight core services and agile light services. Our two-tier deployment strategy ensures robust Quality of Service (QoS) under resource contention. Specifically, core services are placed statically via a long-term network-aware integer program with sparsity constraints to form a fault-tolerant backbone. On the other hand, light services are orchestrated dynamically by a low-complexity online controller that integrates effective capacity theory with \textit{Lyapunov} optimization, providing probabilistic latency guarantees under real-time workload fluctuations. Simulations demonstrate that our framework achieves over 84\% average on-time task completion with moderate deployment costs and maintains strong robustness as the system load scales.

\end{abstract}

\section{Introduction}
The recent advancements of foundation models (FMs) bring a significant shift from pattern-matching to general intelligence. Trained over massive and diverse datasets, FMs, particularly multi-modal FMs, are adept at various complex downstream inference tasks, such as movie generation and real-time augmented reality, by integrating distinct data modalities (e.g., video, text, image, and acoustic signals) \cite{10558825}. Nevertheless, mainstream FM-based services are typically deployed in centralized cloud infrastructure, a setup that inherently suffers from significant bandwidth consumption and potential privacy concerns. Migrating the deployment of FMs to the network edge stands out as a promising solution to enable low-latency and privacy-preserving mobile services\cite{ZW-2025}.

However, deploying FMs on individual edge devices via conventional monolithic architectures is impractical, as the massive internal neurons (or model parameters) demand significant computation and storage resources.
To overcome this limitation and leverage the distributed resources at the network edge, the microservice (MS) architecture is utilized for collaborative edge deployment by functionally decomposing the FM into a collection of small, loosely coupled services \cite{IN-2016}, with computationally lightweight and heavyweight MSs deployed onto edge devices and edge servers, respectively. Consequently, the FM inference task can be modeled as a Directed Acyclic Graph (DAG), where dispersed MSs are interconnected via network links. However, the resulting distributed architecture introduces communication overheads and complex service dependencies, motivating recent optimization and graph-learning approaches for dependency-aware deployment that balance QoS assurance and cost efficiency \cite{CW-2024,WL-2022,WL-2024}.

While these studies address structural complexities, the fluctuating demands from geographically-dispersed users pose a significant threat to operational stability. To circumvent this, a robust adversarial reinforcement learning framework was employed in \cite{ZY-2025} to counteract box uncertainty in task arrivals.
Meanwhile, such external uncertainty is often compounded by internal resource contention. While potentially cost-effective, the pragmatic choice to co-locate multiple services on finite, heterogeneous nodes can lead to severe queuing delays and significant tail latency \cite{SL-2022}. Existing approaches attempt to model this contention either explicitly via predefined average-based conflict coefficients \cite{WL-2024}, which performs poorly for latency-sensitive systems due to heavy tail latency, or implicitly through learning-based resource managers \cite{KF-2021}. Despite these efforts, many works treat microservices as homogeneous components with similar requirements, overlooking their distinct workload and operational characteristics, which, however, are essential for fine-grained resource match, QoS delivery, and efficient scaling in FM-based inference tasks.

Specifically, FM inference pipelines comprise heterogeneous MSs with fundamentally different deployment behaviors. Core MSs, which encapsulate the computation-intensive models (e.g., transformers, vision backbones), exhibit long startup times and limited fault tolerance, making them rigid yet performance-critical anchors of the inference workflow. Conversely, light MSs perform auxiliary, typically stateless operations such as pre- or post-processing, forming an elastic tier that can be rapidly instantiated and parallelized across shared resources to sustain data flow toward the cores. This operational asymmetry creates a multi-timescale coordination challenge, where slow-to-deploy, fault-sensitive core components must coexist with agile, contention-prone light components, demanding a type-aware deployment framework for reliable, responsive, and cost-efficient FM inference.

This work presents a two-tier FM inference framework that leverages the core–light MS dichotomy to manage system uncertainty, with three key contributions:

\begin{enumerate}
\item[$\bullet$] For the first time, we design a hybrid MS deployment strategy for FM edge inference: Slow-startup core services are statically placed to form a reliable system backbone based on long-term workload statistics, while lightweight services are dynamically deployed to elastically adapt to real-time system fluctuations.
\item[$\bullet$] To achieve a forward-looking and fault-tolerant static core MS placement, we formulate a sparsity-constrained integer program that co-optimizes for cost and a statistical QoS score while ensuring deployment diversity.
\item[$\bullet$] To enable QoS-aware online decision-making under resource contention, we pioneer the use of effective capacity theory to link service parallelism with statistical latency bounds, which is then integrated into a \textit{Lyapunov} optimization framework to derive a low-complexity online algorithm for light MS deployment.
\end{enumerate}

\begin{figure*}[t]
    \centering
    \includegraphics[width=0.95\linewidth]{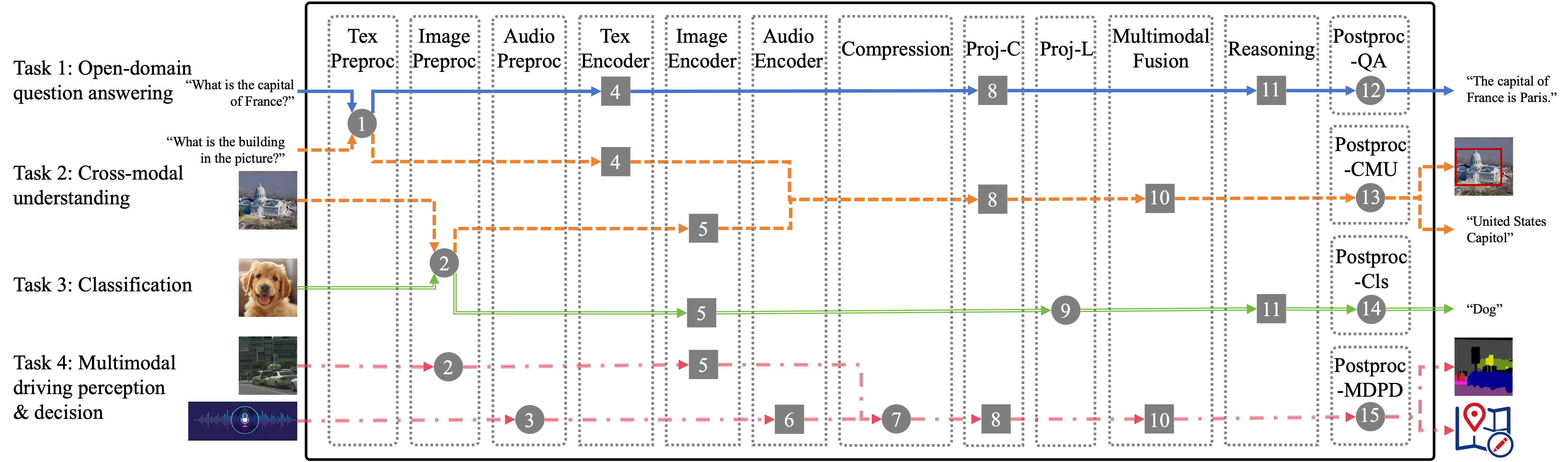}
    \caption{Illustration of FM-based inference application with the MS architecture. Squares denote core MSs, circles denote light MSs, and different line styles represent different task types of inter-service dependencies.}
    \label{fig:MSA} \vspace{-1em}
\end{figure*}

\section{System Model and Problem Formulation}

In this section, we develop an MS-based FM inference framework at the edge network, 
where the FMs are decomposed into a collection of functionally distinct core and light MSs deployed across the network.
We consider a heterogeneous edge network with varying resource capacities, which consists of edge devices (EDs) and edge servers (ESs) represented by $\mathcal{V}$ and interconnected by the set of communication links, $\mathcal{E}$. 
Specifically, resource capacity of node $v$ is denoted by $\mathbf{R}_v=[R_{v,k}]_{k\in[K]}$, where $K$ is the number of distinct resource types (e.g., CPU, RAM, GPU), and $[K]=\{1,2,...,K\}$ is the set of integers up to any $K\in \mathbb{Z}_{>0}$.

\subsection{Microservice Specification for FM Inference}
\begin{figure}[t]
    \centering
    \includegraphics[width=\linewidth]{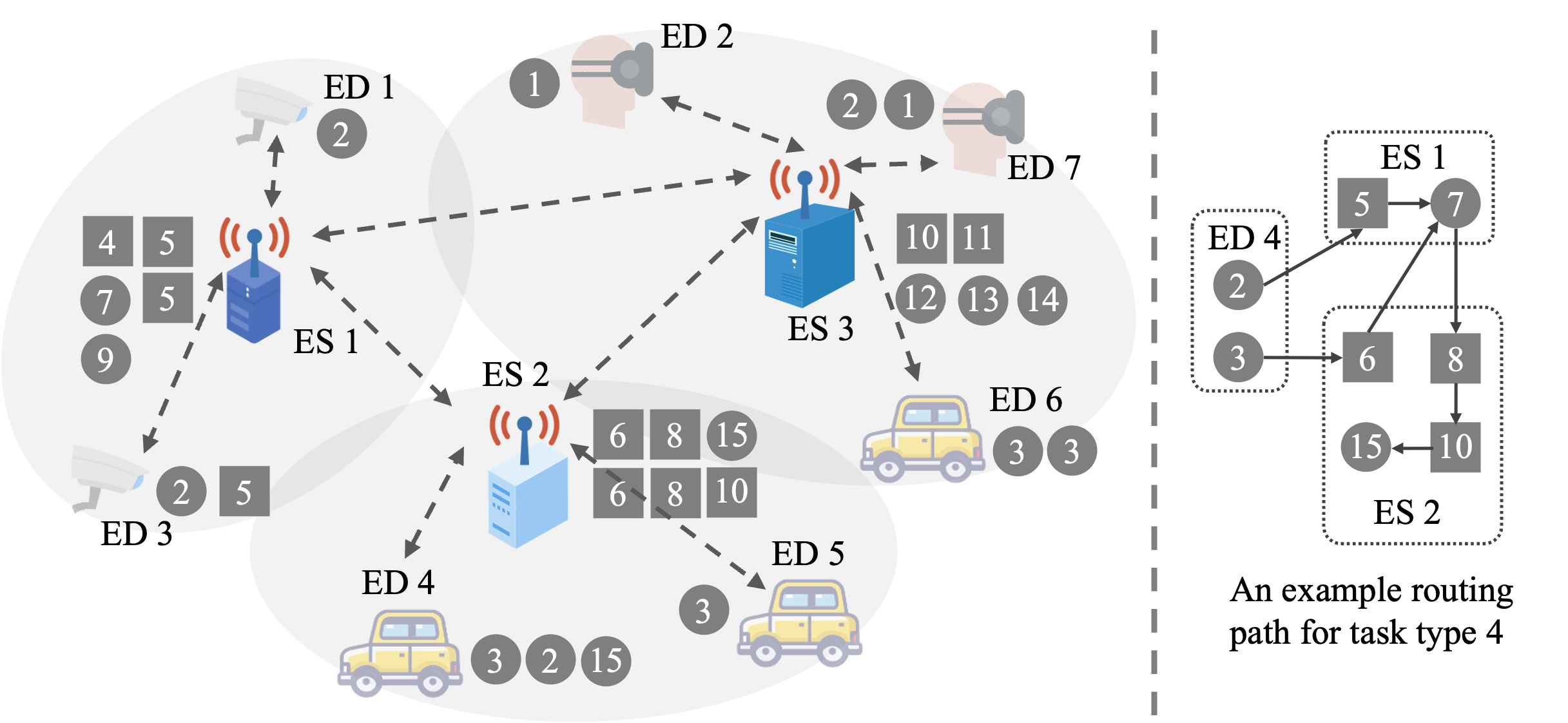}
    \caption{MS-based FM inference on a heterogeneous edge network.}
    \label{fig:NetW}
\end{figure}

An FM inference application architecture is characterized by a fundamental functional asymmetry, which segregates MSs into two categories with distinct operational profiles, as illustrated in Fig. \ref{fig:NetW}. 
Specifically, core MSs ($\mathcal{M}^{\text{cr}}$) are heavyweight and stateful services operated under strict resource isolation to yield deterministic performance, which form the computational backbone. 
Light MSs ($\mathcal{M}^{\text{lt}}$), in contrast, are stateless components with small footprints and can be quickly redeployed elsewhere; their performance is stochastic, a direct result of resource contention incurred by their efficient parallel processing of concurrent tasks on shared resources.

Formally, we characterize each MS $m\in\mathcal{M}=\mathcal{M}^{\text{lt}}\cup \mathcal{M}^{\text{cr}}$ by its resource requirement vector $\mathbf{r}_m=[r_{m,k}]_{k\in[K]}$, and the computational workload, $a_m$ (bits), that must be executed to produce an output of size $b_m$ (bits). Its processing rate, $f_m$ (bits/ms), is a deterministic constant for a core MS but a random variable for any light MS to capture the effects of resource contention. These MSs are orchestrated to execute inference tasks. A task of type $n$ is represented by a DAG $\mathcal{G}_n=(\mathcal{M}_n,\mathcal{L}_n)$, where $\mathcal{M}_n \subseteq \mathcal{M}$ denotes the required MSs ($|\mathcal{M}_n|=I_n$) and $\mathcal{L}_n$ the date dependencies. Consistent with multimodal data fusion, these graphs typically form inverse-tree structures, where each node may have multiple incoming but at most one outgoing edge. starts with an input payload of $A_n$ and must meet an end-to-end latency constraint $D_n$.

\subsection{Latency Formulation under Network Uncertainty}
The system dynamics are driven by several stochastic events. Users $u\in \mathcal{U}$ stochastically generate tasks of type $n$, and we denote the number of such arrivals at time $t$ by the random variable $z_{t,u,n}$. These tasks are transmitted to an associated edge device over a wireless fading channel, where the signal-to-noise ratio $\gamma_u$ is also random. These external uncertainties, along with the processing rates of light MS $f_m$, are assumed to be stationary and statistically independent over time, and their distributions can be accurately profiled.

An admitted task $j$, uniquely identified by its origin $(u,n,t)$, joins the set of active tasks $J(t)$ and is executed along a routing path $P_j=[(v_i,m_i)]_{i\in[I_n]}$ determined by our strategy. Its end-to-end latency comprises the uplink transmission delay from user $u$ to its first node, the inter-node transmission and propagation delay along the routing path, and the processing delay of each invoked service, respectively expressed as
\begin{equation}
    \tau_j^\text{ul} = \frac{A_n}{b_u \log(1+\gamma_u)},
\end{equation}
\begin{equation}
    \tau_j^\text{tr}(v_{i1},v_{i2}) = \frac{b_{m_{i1}}}{w_{(i1,i2)}},\quad \tau_j^\text{pp}(v_{i1},v_{i2})= \frac{W_{(i1,i2)}}{l},
\end{equation}
\begin{equation}
    \tau_j^\text{pc}(v_i)=\frac{a_{m_i}}{f_{m_i}}.
\end{equation}
Here, $b_u$ denotes the bandwidth allocated to user $u$, and $b_u \log(1+\gamma_u)$ is its achievable uplink rate; $w_{(i1,i2)}$ and $W_{(i1,i2)}$ are the bandwidth and distance of the link between nodes $v_{i1}$ and $v_{i2}$; $l$ is the propagation speed.

The DAG dependencies necessitate a recursive calculation for the completion time $T_j$ at any node $v$, as a service must wait for all its predecessors:
\begin{align}
    T_j(v_1)&=\tau_j^\text{ul}+\tau_j^\text{pc}(v_{1}),v_1\in P_j, \nonumber \\
    T_j(v_{i2}) &= \max_{\substack{v_{i1},v_{i2}\in P_j\\v_{i1}\in \mathcal{V}_{P_j}^\text{pa}(v_{i2})}} \left\{T_j(v_{i1})+ \tau_j^\text{tr}(v_{i1},v_{i2})\right. \label{eq:latency} \\[-1.5em]
    &\qquad \qquad \qquad \quad \left. +\tau_j^\text{pp}(v_{i1},v_{i2})+\tau_j^\text{pc}(v_{i2})\right\}, \nonumber
\end{align}
where $\mathcal{V}_{P_j}^\text{pa}(v)$ denotes the set of parent nodes of $v$ in path $P_j$. The total end-to-end latency of task $j$ is
\begin{equation}
    T_j^\text{E2E}=T_j(v_{I_n}),v_{I_n}\in P_j.
\end{equation}
Crucially, $T_j^\text{E2E}$ is a complex, stochastic function of the path $j$, making it challenging to satisfy the deadline $D_n$ while simultaneously optimizing for resource costs.

\subsection{Problem Formulation}
The deployment of core MSs remains fixed throughout a finite time horizon $\mathcal{T}$, governed by $\mathbf{X}^\text{cr}\in\mathbb{N}^{|\mathcal{V}|\times|\mathcal{M}^\text{cr}|}$, where $x_{v,m}^\text{cr}$ denotes the instance number of core MS $m$ placed on node $v$. In contrast, light MSs are deployed dynamically, controlled by two time-varying tensors: $\mathbf{X}^\text{lt} \in \mathbb{N}^{|\mathcal{V}|\times|\mathcal{M}^\text{lt}|\times |\mathcal{T}|}$ specifies the instance count $x_{v,m,t}^\text{lt}$, while $\mathbf{Y}\in \mathbb{N}^{|\mathcal{V}|\times |\mathcal{M^\text{lr}}|\times|\mathcal{T}|}$ defines the parallelism level $y_{v,m,t}$, the number of concurrent tasks an instance can process, to manage resource contention.

The objective is to minimize the total system cost over $\mathcal{T}$. The cost of core MSs includes initial deployment and ongoing maintenance for each instance:
\begin{equation}
    C^\text{cr}(\mathbf{X^\text{cr}})=\sum_{v\in\mathcal{V}}\sum_{m\in\mathcal{M}^\text{cr}} \left( c_m^\text{cr,dp}+\sum_{t\in\mathcal{T}}c_m^\text{cr,mt} \right) x_{v,m}^\text{cr},
\end{equation}
where $c_m^\text{cr,dp}$, $c_m^\text{cr,mt}$ are the one-time deployment price and per-slot maintenance price, respectively. For light MSs, the cost accounts for instantiation, maintenance, and parallelism:
\begin{equation}
    \begin{aligned}
        C^\text{lt}(\mathbf{X}^\text{lt})=&\sum_{v\in\mathcal{V}}\sum_{m\in\mathcal{M}^\text{lt}} \sum_{\substack{t\in\mathcal{T}\\t\neq0}}c_m^\text{lt,dp}\max\{0, x_{t,v,m}^\text{lt} -x_{t-1,v,m}^\text{lt}\}\\
        &+\sum_{v\in\mathcal{V}}\sum_{m\in\mathcal{M}^\text{lt}} \sum_{t\in\mathcal{T}}(c_m^\text{lt,mt}+c_m^\text{lt,pl}) x_{v,m}^\text{lt},
    \end{aligned}
\end{equation}
where $c_m^\text{lt,dp}$, $c_m^\text{lt,mt}$, and $c_m^\text{lt,pl}$ are the instantiation, per-slot maintenance, and parallelism cost of light MS $m$, respectively. 

This optimization is subject to several operational constraints. Firstly, the total resource consumption at any node cannot exceed its capacity:
\begin{equation}
    \begin{aligned}
        \sum_{m\in\mathcal{M}^\text{cr}} r_{m,k}x_{v,m}^\text{cr}+\sum_{m\in\mathcal{M}^\text{lt}} r_{m,k}x_{v,m,t}^\text{lt} \leq R_{v,k},\\
        \forall k\in[K],v\in\mathcal{V},t\in\mathcal{T}.
    \end{aligned}
    \label{resouce_cons}
\end{equation}
To ensure timeliness, the end-to-end latency of every task must meet its deadline:
\begin{equation}
    T_j^\text{E2E} \leq D_n, \forall j\in J(t),t\in\mathcal{T}.
    \label{QoS_cons}
\end{equation}
Furthermore, the provisioned capacity must be sufficient to handle the incoming workload. Let $z_{v,m,t}$ denote the number of tasks requiring MS $m$ at node $v$ at time $t$, which is determined by the routing paths of all concurrent tasks. The deployment must adhere to the following capacity constraints:
\begin{equation}
    x_{v,m}^\text{cr}\geq z_{v,m,t}, \forall m\in \mathcal{M}^\text{cr},v\in\mathcal{V},t\in\mathcal{T},
    \label{capacity_cons1}
\end{equation} 
\begin{equation}
    x_{v,m,t}^\text{lt} y_{v,m,t}\geq z_{v,m,t},\forall m\in \mathcal{M}^\text{lt},v\in\mathcal{V},t\in\mathcal{T}.
    \label{capacity_cons2}
\end{equation}
Finally, all decision variables must be non-negative integers:
\begin{equation}
    x_{v,m}^\text{cr}, x_{v,m,t}^\text{lt}, y_{v,m,t} \in \mathbb{N}, \forall v\in\mathcal{V},m\in\mathcal{M}, t\in\mathcal{T}.
    \label{eq:var_cons}
\end{equation}

The overall MS deployment problem can be formulated as
\begin{equation}
    \begin{aligned}
        \min_{\mathbf{X}^\text{cr},\mathbf{X}^\text{lt},\mathbf{Y}} \quad &C^\text{cr}+C^\text{lt},\\
        \text{s.t.} \quad& \eqref{resouce_cons} \sim \eqref{eq:var_cons}.
    \end{aligned}
    \label{eq:P1}
\end{equation}
Problem \eqref{eq:P1} is intractable due to a triad of compounding challenges. 
First, stochasticity in arrivals and processing rates makes the hard QoS constraint \eqref{QoS_cons} analytically unmanageable, as feasibility itself becomes probabilistic. Second, the workload $z_{v,m,t}$ creates a circular dependency: the optimal deployment strategy depends on the anticipated task load at each node, yet this load itself is a direct consequence of how tasks are routed through the deployed services.
Finally, the problem is high dimensional integer nonlinear programming and renders exact solution methods computationally prohibitive, challenging to meet real-time environments.

\section{Network-Aware Microservice Deployment for FM Edge Inference}
To tackle the challenging problem \eqref{eq:P1}, we propose a two-tier deployment strategy that exploits the functional asymmetry of the FM inference pipelines. 
\subsection{Reliable Core MS Deployment}
The static core MS placement must be cost-efficient and also account for future QoS demands without knowledge of exact, real-time task arrivals. To achieve this, we introduce a heuristic QoS score $Q_{v,m}$ to quantify the expected value of placing an instance of type $m$ on node $v$. By incorporating this score into the objective, we can transform the stochastic deployment into a deterministic integer program:
\begin{equation}
    \begin{aligned}
        \min_{\mathbf{X}^\text{cr}} \quad & \sum_{v\in\mathcal{V}}\sum_{m\in \mathcal{M}^\text{cr}} x_{v,m}^\text{cr}(c_m^\text{cr}-\xi Q_{v,m}), \\
        \text{s.t.}\quad & \text{C1: } r_{m,k}x_{v,m}^\text{cr}\leq R_{v,k}, \forall k\in[K],v\in\mathcal{V},\\
        & \text{C2: } \sum_{v\in\mathcal{V}} \Tilde{z}_{v,m}\leq \sum_{v\in\mathcal{V}} x_{v,m}^\text{cr}, \forall m\in\mathcal{M}^\text{cr},\\
        & \text{C3: } x_{v,m}^\text{cr}\in \mathbb{N}, \forall v\in\mathcal{V},m\in \mathcal{M}^\text{cr},
    \end{aligned}
    \label{eq:P2}
\end{equation}
where, $c_m^\text{cr}=c_m^\text{cr,dp}+c_m^\text{cr,mt}$ and $\xi\geq0$ is a weight balancing cost against the QoS measure. The term $\Tilde{z}_{v,m}$ is the average estimate for load $z_{v,m,t}$. For this static formulation, the real-time per-node capacity constraint \eqref{capacity_cons1} is relaxed into a global  constraint, which ensures the total long-term capacity meets the total estimated demand across the network.

The heuristics $\Tilde{z}_{v,m}$ and $Q_{v,m}$ are derived from a mean-value analysis of latency profiles. To obtain these estimates, we consider a typical task $j$ (identified by its origin $(u,n,t)$) requiring MS $m\in \mathcal{M}^\text{cr}\cap \mathcal{M}_n$ at node $v\in\mathcal{V}$ and partition its estimated end-to-end latency into three parts: the preceding latency to reach node $v$, $d_j^\text{pr}(v,m)=\max_{v^{'}\in\mathcal{V}_{P_{j,v}}^\text{pa}(v)} \bar{T}_j(v^{'})$; the processing time at the current node, $d_j^\text{cu}(v,m)=\frac{a_m}{f_m}$; and the succeeding latency for all subsequent MSs, $d_j^\text{su}(v,m)=\sum_{m^{'}\in\mathcal{M}_n^\text{de}(m)} \frac{a_{m^{'}}}{\bar{f}_{m^{'}}}$, respectively.
Specifically, $\bar{T}_j$ is calculated via \eqref{eq:latency} using mean values for all random variables, $P_{j,v}$ is the shortest path from the task's source to node $v$ where path length is measured as the sum of network and average computation latencies, and $\mathcal{M}_n^\text{de}(m)$ is the set of descendant MSs of $m$ in $\mathcal{G}_n$. 

For the estimated load $\Tilde{z}_{v,m}$, we apportion the mean task arrival rate to nodes based on an exponential decay of the preceding latency:
\begin{equation}
    \Tilde{z}_{v,m} = \sum_{u\in\mathcal{U}}\sum_{n\in\mathcal{N}_m} \frac{e^{-\delta d_j^\text{pr}(v,m)}}{\sum_{v^{'}\in\mathcal{V}}e^{-\delta d_j^\text{pr}(v^{'},m)}} \mathbb{E}[z_{u,n,t}],
\end{equation}
where $\mathcal{N}_m$ is the set of task type requiring MS $m$. This model allocates higher load to nodes strategically closer to users. The QoS score captures placement urgency and is the product of the estimated load and an ``urgency metric," $\Tilde{d}_{v,m}$, which quantifies timeliness feasibility with the capped ratio of the remaining time budget to the estimated future processing time:
\begin{equation*}
\begin{split}
    &\Tilde{d}_{v,m}=\sum_{u\in\mathcal{U}}\sum_{n\in\mathcal{N}}\max\left\{\frac{D_n-d_j^\text{pr}(v,m)-d_j(v,m)}{d_j^\text{su}(v,m)},C_1\right\},\\
    &Q_{v,m}=\Tilde{z}_{v,m}\Tilde{d}_{v,m},
\end{split}
\end{equation*}
where $C_1$ is a constant. A high QoS score thus signifies a strategically valuable placement expected to serve a high volume of tasks that can comfortably meet their deadlines.

While \eqref{eq:P2} is a standard integer program solvable by off-the-shelf tools, solvers tend to yield sparse solutions that consolidate all instances of an MS onto a single node and lead to a single-point vulnerability. To enhance deployment diversity, we introduce a binary auxiliary variable $\hat{x}_{v,m}$ for each $x_{v,m}^\text{cr}$ and add the following constraints:
\begin{equation}
    \begin{aligned}
        &\text{C4: }  x_{v,m}^\text{cr} \leq C_2 \hat{x}_{v,m}, \forall v\in\mathcal{V},m\in \mathcal{M}^\text{cr},\\
        & \text{C5: } x_{v,m}^\text{cr} \geq C_3 \hat{x}_{v,m}, \forall v\in\mathcal{V},m\in \mathcal{M}^\text{cr}, \\
        & \text{C6: }\sum_{v\in\mathcal{V}}\sum_{m\in \mathcal{M}^\text{cr}}\hat{x}_{v,m}\geq \kappa,
    \end{aligned}
\end{equation}
where $C_2$ and $C_3$ are sufficiently large and small positive constants, respectively. C4 and C5 jointly force $\hat{x}_{v,m}=1$ if and only if $x_{v,m}^\text{cr}>0$. C6 imposes a minimum of $\kappa$ non-zero deployments across the network, thereby preventing over-centralization. The parameter $\kappa$ tunes the trade-off between the objective value of \eqref{eq:P2} and system reliability.

\subsection{Dynamic Light MS Deployment}
With the static placement of core MSs fixed, the remaining node capacities are dedicated to the dynamic deployment of light MSs, which constitutes an online stochastic optimization problem formulated as:
\begin{equation}
    \begin{aligned}
        \min_{\mathbf{X}^\text{lt}_t,\mathbf{Y}_t} \quad &C^\text{lt}, \\
        \text{s.t.} \quad &\text{C1: } \sum_{m\in\mathcal{M}^\text{lt}} r_{m,k}x_{v,m,t}^\text{lt}\leq R_{v,k}^\text{lt}, \forall v\in\mathcal{V}, k\in[K],\\
        &\text{C2: } T_j^\text{E2E}\leq D_n,\forall j\in J(t), t\in \mathcal{T}, \\
        &\text{C3: } x_{v,m,t}^\text{lt}y_{v,m,t}\leq z_{v,m,t}, \forall v\in\mathcal{V},m\in\mathcal{M}^\text{lt},t\in\mathcal{T}, \\
        &\text{C4: } x_{v,m,t}^\text{lt}, y_{v,m,t} \in \mathbb{N}, \forall v\in\mathcal{V},m\in\mathcal{M}^\text{lt},t\in\mathcal{T},
    \end{aligned}
    \label{eq:P3}
\end{equation}
where $R_{v,k}^\text{lt}=R_{v,k}-\sum_{m\in\mathcal{M}^\text{cr}} r_{m,k}x_{v,m}^\text{cr}$. The instance counts $\mathbf{X}^\text{lt}_t$ and parallelism levels $\mathbf{Y}_t$ must be decided at each slot based solely on the current system state, without knowledge of future arrivals, to optimize long-term cost.

To address \eqref{eq:P3}, we employ the \textit{Lyapunov} drift-plus-penalty framework \cite{10855336}, a powerful tool for online stochastic control. The core idea is to transform C2 into a queue stability condition. Specifically, a virtual queue $H_j(t)$ is introduced for each task $j$ to track its cumulative deadline violations:
\begin{equation}
    H_j(t+1)=\max\left\{H_j(t)+T_j(t)- D_n,\zeta\right\},\forall j\in J(t),
\end{equation}
where $T_j(t)$ is latency experienced by task $j$ under the decisions made by time $t$. Different from \cite{10855336}, we introduce a floor $\zeta>0$ to prevent the virtual queue from collapsing to zero, thereby keeping the controller proactively incentivized to maintain low latency rather than reacting only once latency has already accumulated. The framework then seeks to minimize the following drift-plus-penalty expression at each slot, subject to instantaneous operational constraints C1, C3, and C4:
\begin{equation}
    L=\eta C^\text{lt}+\sum_{j\in J(t)} \phi_j H_j(t)[T_j(t)- D_n], v_i\in P_j,
    \label{eq:subobj}
\end{equation}
where $\eta\geq0$ is a tunable parameter controlling cost-latency trade-off, and $\phi_j$ is a task-specific priority weight. 

A key challenge in minimizing \eqref{eq:subobj} lies in the stochastic nature of the latency term $T_j(v_i)$, which depends nonlinearly on the deployment decisions $\{\mathbf{X}^\text{lt}_t,\mathbf{Y}_t\}$. While network latency is deterministic once a route is chosen, the processing times are uncertain due to resource contention. We adopt the effective capacity model \cite{MA-2019} to bridge this gap. Rooted in large deviations theory, this framework analytically characterizes the maximum constant workload arrival rate that a stochastic server can sustain under a prescribed latency violation probability. The effective capacity of a light MS $m$ is given by:
\begin{equation}
    E^\text{c}_m(\theta)=-\lim_{t\rightarrow \infty}\frac{\ln \mathbb{E}[e^{-\theta F_m(0,t)}]}{\theta t},
    \label{eq:ec}
\end{equation}
where $F(0,t)=\sum_{\tau = 0}^{t-1}f_m(t)$ is the cumulative service process. The QoS exponent $\theta>0$ directly links the effective capacity to the tail probability of the latency distribution:
\begin{equation}
    \mathbb{P}\{d>D\}\approx\frac{E^\text{c}_m(\theta)}{\mathbb{E}[f_m(t)]}e^{-\theta E^\text{c}_m(\theta)D}. 
    \label{eq:lvp}
\end{equation}
This relationship enables the pre-calculation of a deterministic mapping $d=g_{m,\epsilon}(y)$ between the chosen parallelism level $y$ and the resulting processing time $d$ that satisfies the violation probability $\epsilon$, thereby facilitating efficient online adaptation.

Even with this transformation, the per-slot optimization problem remains a complex non-linear integer program that cannot be solved optimally within the tight time constraints of online operation. We therefore propose a low-complexity greedy heuristic, detailed in Algorithm \ref{alg:light_deploy}, which iteratively makes the single deployment decision that yields the largest marginal decrease in \eqref{eq:subobj}. 

At the beginning of each time slot, we first determine the set of busy instances from the previous slot, $x_{v,m,t-1}^{\text{lt,bs}}$, which are still processing ongoing tasks. Then, in a greedy loop, for all tasks in the waiting queue $J^\text{qu}(t)$, we calculate the potential change in the objective, $\Delta_{v,m}L$, for every feasible incremental deployment (adding one instance of $m$ on node $v$). This involves routing each task to the instance that minimizes its next-hop latency, which is the sum of network delay and the QoS-aware processing delay from our mapping function:
\begin{equation*}
    \Delta T_j(v,m)=\tau_j^\text{tr}(v_{j},v)+\tau_j^\text{pp}(v_{j},v)+g_{m,\epsilon}(y_{v,m,t}+1)(v),
\end{equation*}
where $v_{j}$ the node currently hosting the last completed service for task $j$.
The algorithm then implements the deployment with the most negative $\Delta_{v,m}L$, updates the system state, and repeats until no further cost-effective deployments can be made. Assuming the greedy selection loop runs $M$ times, the per-slot complexity is $\mathcal{O}(M(1+|J^\text{qu}(t)||\mathcal{V}||\mathcal{M}^\text{lt}|))$.

\begin{algorithm}
\caption{Greedy Online Light MS Deployment}
\label{alg:light_deploy}
\For{each $t \in \mathcal{T}$}{
    $x_{v,m,t}^{\text{lt}} \leftarrow x_{v,m,t-1}^{\text{lt,bs}}$ and $y_{v,m,t}\leftarrow 0$ for all $(v,m)$;

    Observe queue $H_j(t)$ for all $j\in J^\text{qu}(t)$;

    \While{True}{
        $(v^\ast,m^\ast) \leftarrow \argmin\limits_{v\in\mathcal{V},\,m\in\mathcal{M}^{\text{lt}}} \Delta_{v,m}L$;\\

        \If{$\Delta_{v^\ast,m^\ast}L < 0$}{
            $x_{v^\ast,m^\ast,t}^{\text{lt}} \leftarrow x_{v^\ast,m^\ast,t}^{\text{lt}} + 1$; 
        }
        \ElseIf{$\Delta_{v^\ast,m^\ast}L = \infty$}{
            \textbf{break};
        }
        \Else{
             $\Delta_{v^\ast,m^\ast}L \leftarrow \infty$;
        }

        \eIf{$R_{v^\ast,k}(t) > r_{m^\ast,k},\ \forall k \in [K]$}{
            $\Delta_{v^\ast,m^\ast}L \leftarrow \eta c_{m^\ast}^{\text{lt}}$;\\
            \For{each $j \in J^{\text{qu}}(t)$}{
                $(v_j,m_j) \leftarrow \argmin\limits_{v\in\mathcal{V},\,m\in\mathcal{M}^{\text{lt}}} \Delta T_j(v,m)$;\\
                $y_{v_j,m_j,t} \leftarrow y_{v_j,m_j,t} + 1$;\\
                $\Delta_{v^\ast,m^\ast}L \leftarrow \Delta_{v^\ast,m^\ast}L + \phi_j H_j(t)\Delta T_j(v_j,m_j)$;
            }
        }{
            $\Delta_{v^\ast,m^\ast}L \leftarrow \infty$;
        }
    }
}
\end{algorithm}

\section{Performance Evaluation}
We evaluate an FM-based edge inference application comprising 4 task types, 6 core MSs, and 9 light MSs, with dependencies shown in Fig. \ref{fig:MSA}, deployed on an edge network (Fig. \ref{fig:NetW}). Key parameters are listed in TABLE \ref{table:para}; values for each run are sampled from predefined ranges. Our proposal derives the processing latency of light MSs from the effective capacity models configured with a violation probability $\epsilon=0.2$.

We evaluate our proposal against three baseline methods:
\begin{itemize}
    \item \textbf{LBRR}: Services are allocated to the least-loaded nodes. Incoming tasks are then scheduled across available instances using a Round-Robin policy.
    \item \textbf{GA}: A metaheuristic optimizer that seeks a near-optimal deployment by minimizing a fitness function combining total system cost and QoS violation rates.
    \item \textbf{PropAvg}: A direct ablation of our framework. It employs the same two-tier logic as our proposal but replaces the Effective Capacity model with a simpler mean-value estimation for processing delays.
\end{itemize}
{\renewcommand{\arraystretch}{1.1} 
\begin{table*}[t]
\centering
\caption{Key Simulation Parameters}
\begin{tabular}{c c c c c c c}
\toprule
 & $r_{m,k}/R_{v,k}$(CPU;RAM;GPU;VRAM) & $a_m$(MB) & $b_m$(MB) & $f_m$(MB/ms) & 
 \multicolumn{2}{c}{$c_m^\text{dp},c_m^\text{mt},c_m^\text{pl}$}\\ 
\hline
\textbf{Core MS} & [2,16];[1,4];[4,32];[4,32] & [2,16] & [0.1,1] & [8,32] & \multicolumn{2}{c}{20.0; 4.0; 0.0}  \\ 

\textbf{Light MS} & [0.5,2];[0,0.5];[0.25,4];[0,1] & [0.5,2] & [0.25,1.5] & $Gamma$([1,2],[1,20])  & \multicolumn{2}{c}{4.0; 1.0; 0.5} \\ 
\cline{3-7}
\textbf{ED} & [1,64];[1,32];[0,64];[0,64] & \multicolumn{1}{c}{$z_{u,n,t}$(/ms)} & $D_n$(ms) & $\gamma_u$(Gbs) &  $A_n$(MB) & $w$(MB/ms) \\
\cline{3-7}
\textbf{ES} & [128,256];[64,128];[1024,2048];[256,512] & \multicolumn{1}{c}{$Poisson$([0.15,1.5])} & [50,100] & $Nakagami$([1.5,3],[0.5,1]) & [0.5,4] & [0.1,1.0]\\ 
\bottomrule
\end{tabular} \vspace{-1em}
\label{table:para}
\end{table*}}

\begin{figure}[t]
    \centering
    \includegraphics[width=0.9\linewidth]{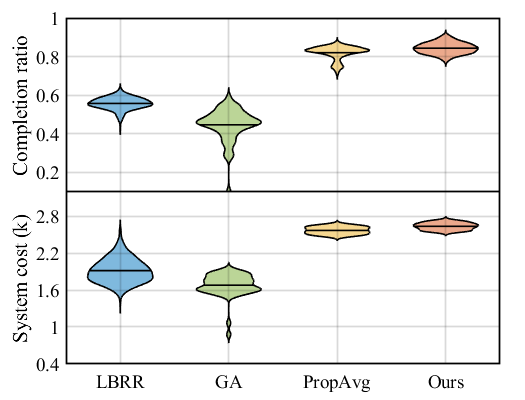}
    \caption{Violin-plot comparison of on-time task completion rate and total system cost across four deployment strategies.}
    \label{fig:Res1}
\end{figure}

\begin{figure}[t]
    \centering \vspace{-1em}
    \includegraphics[width=0.95\linewidth]{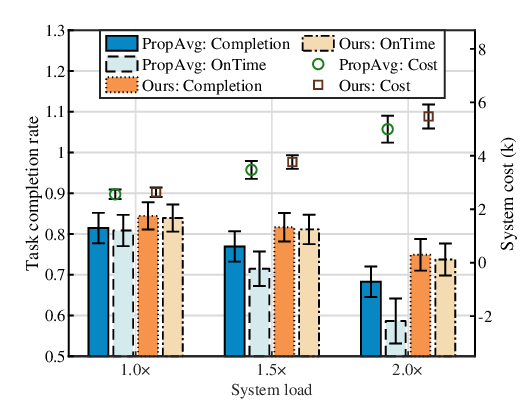}
    \caption{Comparison of the proposed framework and the PropAvg ablation under escalating system loads.}\vspace{-1em}
    \label{fig:Res2}
\end{figure}

Fig. \ref{fig:Res1} visualizes, via violin plots, the distributions of the on-time task completion rate and the total system cost across four deployment strategies. 
Narrow, concentrated violins indicate stable performance, whereas wider ones suggest inconsistency. An effective deployment should exhibit a compact distribution with high completion rates and moderate costs. 
Our proposal exhibits a sharply peaked and compact distribution centered around a high completion rate (above 84\%), indicating consistently reliable QoS across trials. Its cost distribution is similarly concentrated around a justifiable level, demonstrating stable and efficient resource utilization. 
The \textbf{LBRR} method, while simple, yields results in a low-cost, low-performance regime, an expected outcome given its deadline-agnostic nature.
In contrast, \textbf{GA} results exhibit a widely distributed distribution for both metrics, reflecting high variability in the results. This instability suggests that the metaheuristic search struggles to converge in the vast and stochastic optimization space, leading to under-provisioning and inconsistent performance.
The \textbf{PropAvg} ablation variant produces slightly lower costs than our proposal but a broader and skewed completion-rate distribution with a long lower tail. This pattern reveals its core limitation: mean-based estimation fails to capture tail-latency effects, reducing cost marginally at the expense of intermittent but severe QoS degradation.

Fig. \ref{fig:Res2} compares our proposal and the \textbf{PropAvg} ablation under escalating system loads ($1.0\times$, $1.5\times$, $2.0\times$ multipliers to the mean of the task arrival distribution). 
Bars with error bars (left axis) show total and on-time completion rates, and markers (right axis) indicate system cost. Higher and closer completion rates imply stronger QoS resilience, while lower, steadier cost curves indicate better scalability.
As system load increases, both methods show declining completion performance and rising costs, with growing variance and steeper trends. While \textbf{PropAvg} maintains a growing cost advantage, its completion rates rapidly deteriorate as the system becomes saturated. Moreover, the gap between its total and on-time completion rates widens significantly, reflecting frequent deadline violations even for finished tasks. In contrast, our proposal maintains both high completion and on-time rates, with only marginal gap expansion and controlled cost scaling. These results demonstrate that average-based methods like \textbf{PropAvg} are brittle under stress, while our proposal achieves robust, time-consistent QoS and cost-efficient scalability even in heavy-load conditions.

\section{Conclusions}
This paper addressed QoS-aware and cost-efficient MS deployment for FM inference at the resource-contended edge under uncertainties in task arrivals and wireless channels. Our solution is a hybrid static-dynamic framework founded on the intrinsic functional asymmetry between core and light MSs. The static tier provides a reliable computational backbone by co-optimizing long-term operational cost, statistical QoS, and deployment diversity of core services. The dynamic tier complements this with an online control mechanism that governs light services under stochastic conditions, regulating latency violations through effective-capacity-based \textit{Lyapunov} optimization. Numerical evaluations confirmed that the proposed framework consistently achieves superior cost-QoS balance and resilience under intensifying loads. This marks a critical advance for robust edge AI, with future work poised to explore learning-based solvers and accuracy-aware optimization.

\bibliographystyle{IEEEtran}
\bibliography{ref}
\setlength{\itemsep}{-0.5ex}
\end{document}